%% file: nnls_laa.tex
\documentclass[11pt]{article}
\usepackage{algorithm2e,framed}
\usepackage{amssymb,amsfonts,amsmath}
\usepackage{algorithm2e,framed}
\usepackage{amsfonts,amsmath}
\usepackage{graphicx}

\input{newcmds}

\title{\textbf{Random Projections for the Nonnegative Least-Squares Problem}}

\author{
\textbf{Christos }\textbf{Boutsidis}\\
Computer Science Department\\
Rensselaer Polytechnic Institute\\
Troy, NY 12180 \\
\texttt{boutsc@cs.rpi.edu} \\
\and
\textbf{Petros Drineas}\\
Computer Science Department\\
Rensselaer Polytechnic Institute\\
Troy, NY 12180 \\
\texttt{drinep@cs.rpi.edu} \\
}

\begin{document}

\maketitle

\begin{abstract}
Constrained least-squares regression problems, such as the
Nonnegative Least Squares (NNLS) problem, where the variables are
restricted to take only nonnegative values, often arise in
applications. Motivated by the recent development of the fast
Johnson-Lindestrauss transform, we present a fast random
projection type approximation algorithm for the NNLS problem. Our
algorithm employs a randomized Hadamard transform to construct a
much smaller NNLS problem and solves this smaller problem using a
standard NNLS solver. We prove that our approach finds a
nonnegative solution vector that, with high probability, is close
to the optimum nonnegative solution in a relative error
approximation sense. We experimentally evaluate our approach on a
large collection of term-document data and verify that it does
offer considerable speedups without a significant loss in
accuracy. Our analysis is based on a novel random projection type
result that might be of independent interest. In particular, given
a tall and thin matrix $\Phi \in \mathbb{R}^{n \times d}$ ($n \gg
d$)  and a vector $y \in \mathbb{R}^d$, we prove that the
Euclidean length of $\Phi y$ can be estimated very accurately by
the Euclidean length of $\tilde{\Phi}y$, where $\tilde{\Phi}$
consists of a small subset of (appropriately rescaled) rows of
$\Phi$.
\end{abstract}

\section{Introduction}
The Nonnegative Least Squares (NNLS) problem is a constrained
least-squares regression problem where the variables are allowed
to take only nonnegative values. More specifically, the NNLS
problem is defined as follows:
\begin{definition} \label{def:NNLS}
\textbf{\textsc{[Nonnegative Least Squares~(NNLS)]}}\\
Given a matrix $A \in \mathbb{R}^{n \times d}$ and a target vector
$b \in \mathbb{R}^{n}$, find a nonnegative vector $x_{opt} \in
\mathbb{R}^{d}$ such that
\begin{equation}
\label{eqn:def}
 x_{opt} = \arg \min_{x \in \mathbb{R}^d, x\geq 0} \TNormS{Ax -
b}.
\end{equation}
\end{definition}
NNLS is a quadratic optimization problem with linear inequality
constraints. As such, it is a convex optimization problem and thus
it is solvable (up to arbitrary accuracy) in polynomial time
\cite{Bjo96}. In words, NNLS seeks to find the best nonnegative
vector $x_{opt}$ in order to approximately express $b$ as a
strictly nonnegative linear combination of the columns of $A$,
i.e., $b \approx Ax_{opt}$.

The motivation for NNLS problems in data mining and machine
learning stems from the fact that given least-squares regression
problems on nonnegative data such as images, text, etc., it is
natural to seek \emph{nonnegative} solution vectors. (Examples of
data applications are described in \cite{CP07}.) NNLS is also
useful in the computation of the Nonnegative Matrix
Factorization~\cite{KP07a}, which has received considerable
attention in the past few years. Finally, NNLS is the core
optimization problem and the computational bottleneck in designing
a class of Support Vector Machines~\cite{SSL02}. Since modern
datasets are often massive, there is continuous need for faster,
more efficient algorithms for NNLS.

In this paper we discuss the applicability of random projection
algorithms for solving constrained regression problems, and in
particular NNLS problems. Our goal is to provide fast
approximation algorithms as alternatives to the existing exact,
albeit expensive, NNLS methods. We focus on input matrices $A$
that are tall and thin, i.e., $n \gg d$, and we present, analyze,
and experimentally evaluate a random projection type algorithm for
the nonnegative least-squares problem. Our algorithm utilizes a
novel random projection type result which might be of independent
interest. We argue that the proposed algorithm (described in
detail in Section \ref{sec:algo}), provides relative error
approximation guarantees for the NNLS problem. Our work is
motivated by recent progress in the design of fast randomized
approximation algorithms for unconstrained $\ell_p$ regression
problems~\cite{DMMS07,DDHKM08}.

The following theorem is the main
quality-of-approximation result for our randomized NNLS algorithm.
\begin{theorem}
\label{thm:main_result} Let $\epsilon \in (0,1]$. Let $A \in
\mathbb{R}^{n \times d}$ and $b \in \mathbb{R}^{n}$ be the inputs
of the NNLS problem with $n \gg d$. If the input parameter $r$ of
the \textsc{RandomizedNNLS} algorithm of Section~\ref{sec:algo}
satisfies
\begin{equation}
\label{eqn:result0_intro} \frac{r}{\log r} \geq \frac{342 c_o^2
(d+1) \log(n)}{\epsilon^2},
\end{equation}
(for a sufficiently large constant $c_o$)\footnote{$c_o$ is an
unspecified constant in \cite{RV06}.} then the
\textsc{RandomizedNNLS} algorithm returns a nonnegative vector
$\tilde{x}_{opt}$ such that
\begin{equation}
\label{eqn:result1_intro} \TNormS{A\tilde{x}_{opt}-b} \le
(1+\epsilon) \min_{x\in \mathbb{R}^d, x\geq 0}\TNormS{Ax-b},
\end{equation}
holds with probability at least $0.5$\footnote{Note that a small
number of repetitions of the algorithm suffices to boost its
success probability.}. The running time of the
\textsc{RandomizedNNLS} algorithm is
\begin{equation}
\label{eqn:result3_intro}O(n d \log (r)) + T_{NNLS}\left(r,
d\right).
\end{equation}
The latter term corresponds to the time required to exactly solve
an NNLS problem on an input matrix of dimensions $r \times d$.
\end{theorem}
One should compare the running time of our method to
$T_{NNLS}\left(n, d\right)$, which corresponds to the time
required to solve the NNLS problem exactly. We experimentally
evaluate our approach on 3,000 NNLS problems constructed from a
large and sparse term-document data collection. On average (see
section \ref{sec:exppart1}), the proposed algorithm achieves a
three-fold speedup when compared to a state-of-the-art NNLS
solver~\cite{KSD07} with a small (approx. $10 \% $) loss in
accuracy; a two-fold speedup is achieved with a $4\%$ loss in
accuracy. Computational savings are more pronounced for NNLS
problems with denser input matrices $A$ and vectors $b$ (see
section \ref{sec:sparsedense}).

The remainder of the paper is organized as follows. Section
\ref{sec:priorwork} reviews basic linear algebraic definitions and
discusses related work. In Section \ref{sec:algo} we present our
randomized algorithm for approximating the NNLS problem, discuss
its running time, and give the proof of Theorem
\ref{thm:main_result}. Finally, in section \ref{sec:exp} we
provide an experimental evaluation of our method.

\section{Background and related work} \label{sec:priorwork}
Let $[n]$ denote the set $\{1,2,\ldots,n\}$. For any matrix $A \in
\mathbb{R}^{n \times d}$ with $n \geq d$ let $A_{(i)}, i \in [n]$
denote the $i$-th row of $A$ as a row vector, and let $A^{(j)}, j
\in [d]$ denote the $j$-th column of $A$ as a column vector. The
Singular Value Decomposition (SVD) of $A$ can be written as
\begin{align}
\label{svdA} A = U_A \Sigma_A V_A^T.
\end{align}
Assuming that $A$ has full rank, $U_A \in \mathbb{R}^{n \times d}$
and $V_A \in \mathbb{R}^{d \times d}$  are orthonormal matrices,
while $\Sigma_A$ is a $d \times d$ diagonal matrix. Finally, $
\FNormS{A} = \sum_{i=1}^n \sum_{j=1}^d A_{ij}^2 $ denotes the
square of the Frobenius norm of $A$ and $\TNorm{A} = \sup_{x\in
\mathbb{R}^d,\ x\neq 0} \TNorm{Ax}/\TNorm{x}$ denotes the spectral
norm of $A$.

The (non-normalized) $n \times n$ matrix of the Hadamard-Walsh
transform $H_n$ is defined recursively as follows:
$$ H_n = \left[
\begin{array}{cc}
  H_{n/2} & H_{n/2} \\
  H_{n/2} & -H_{n/2}
\end{array}\right]   ,
\qquad \mbox{with} \qquad
H_2 = \left[
\begin{array}{cc}
  +1 & +1 \\
  +1 & -1
\end{array}\right].
$$
The $n \times n$ normalized matrix of the Hadamard-Walsh transform
is equal to $\frac{1}{\sqrt{n}}H_n$; hereafter, we will denote
this normalized matrix by $H_n$ ($n$ is a power of $2$). For
simplicity, throughout this paper we will assume that $n$ is a
power of two; padding $A$ and $b$ with all-zero rows suffices to
remove the assumption. Finally, all logarithms are base two.

\subsection{Random projection algorithms for unconstrained $\ell_{p}$ problems}
The unconstrained least-squares regression problem ($\ell_2$)
takes as input a matrix $A \in \mathbb{R}^{n \times d}$ and a
vector $b \in \mathbb{R}^n$, and returns as output a vector
$x_{opt} \in \mathbb{R}^d$ that minimizes the distance
$\TNormS{Ax-b}$. Assuming $n \gg d$, various algorithms solve the
problem exactly in $O(nd^2)$ time~\cite{GV89}. Drineas et
al.~\cite{DMM06, DMMS07} and Sarlos~\cite{Sar06} give randomized
algorithms to approximate the solution to such problems. The basic
idea of these algorithms is to select a subset of rows from $A$
and a subset of elements from $b$, and solve the induced problem
exactly. The fundamental algorithmic challenge is how to form the
induced problem. It turns out that sampling rows of $A$ and
elements of $b$ with probabilities that are proportional to the
$\ell_2$ norms of the rows of the matrix of the left singular
vectors of $A$ suffices \cite{DMM06}. This approach is not
computationally efficient, since computing these probabilities
takes $O(nd^2)$ time. However, by leveraging the Fast
Johnson-Lindenstrauss Transform of~\cite{AC06}, one can design an
$o(nd^2)$ algorithm for this problem~\cite{DMMS07}. The algorithm
of this paper applies the same preconditioning step as the main
Algorithm of \cite{DMMS07}. The analysis though is very different
from the analysis of \cite{DMMS07} and is based on a novel random
projection type result that is presented in section
\ref{sec:algo}. The difficulty of applying the analysis of
\cite{DMMS07} here is the fact that the solution of an NNLS
problem cannot be written in a closed form. Finally, it should be
noted that similar ideas are discussed in \cite{DDHKM08}, where
the authors present sampling-based approximation algorithms for
the $\ell_p$ regression problem for $p=[1,\infty)$. The
preconditioning step of the algorithm of this paper is different
from the preconditioning step of the algorithm of \cite{DDHKM08}.

\subsection{Algorithms for the NNLS problem}
We briefly review NNLS algorithms following the extensive review
in~\cite{CP07}. Recall that the NNLS Problem is a quadratic
optimization problem. Hence, all quadratic programming algorithms
may be used to solve it. Methods for solving NNLS problems can be
divided into three general categories: ($i$) active set methods,
($ii$) iterative methods, and ($iii$) other methods. The approach
of Lawson and Hanson in \cite{LH74} seems to be the first
technique to solve NNLS problems. It is a typical example of an
active set method and is implemented as the function
\emph{lsqnonneg} in Matlab. Immediate followups to this work
include the technique of Bro and Jong~\cite{BJ97} which is
suitable for problems with multiple right hand sides, as well as
the combinatorial NNLS approach of Dax~\cite{Dax91}. The
Projective Quasi-Newton NNLS algorithm of \cite{KSD07} is an
example from the second category. It is an iterative approach
based on the Newton iteration and the efficient approximation of
the Hessian matrix. Numerical experiments in \cite{KSD07} indicate
that it is a very fast alternative to the aforementioned active
set methods. The sequential coordinate-wise approach
of~\cite{FHN05} is another example of an iterative NNLS method.
Finally, interior point methods are suitable for NNLS
computations~\cite{PJV94}. A different approach appeared in
\cite{SSL02}. It starts with a random nonnegative vector $x \in
\mathbb{R}^d$ and updates it via elementwise multiplicative rules.
Surveys on NNLS algorithms include~\cite{Bjo96, LH74, KSD07}.

\section{A Random Projection Type Algorithm for the NNLS problem}
\label{sec:algo}

This section describes our main algorithm for the NNLS problem.
Our algorithm employs a randomized Hadamard transform to construct
a much smaller NNLS problem and solves this smaller problem
exactly using a standard NNLS solver. The approximation accuracy
of our algorithm is a function of the size of the small NNLS
problem.

\subsection{The \textsc{RandomizedNNLS} Algorithm}
\label{sec:RNNLS}

Algorithm \textsc{RandomizedNNLS} takes as inputs an $n \times d$
matrix $A$ ($n \gg d$), an $n$-dimensional target vector $b$, and
a positive integer $r < n$. It outputs a nonnegative
$d$-dimensional vector $\tilde{x}_{opt}$ that approximately solves
the original NNLS problem. Our algorithm starts by premultiplying
the matrix $A$ and the right hand side vector $b$ with a random $n
\times n$ diagonal matrix $D$, whose diagonal entries are set to
$+1$ or $-1$ with equal probability. It then multiplies the
resulting matrix $DA$ and the vector $Db$ with a small submatrix
of the $n \times n$ normalized Hadamard-Walsh matrix $H_n$ (see
section \ref{sec:priorwork}). This submatrix of $H_n$ -- denoted
by $\tilde{H}$ -- is constructed as follows: for all $i \in [n]$,
the $i$-th row of $H_n$ is included in $\tilde{H}$ with
probability $r/n$. Clearly, the expected number of rows of the
matrix $\tilde{H}$ is equal to $r$. Finally, our algorithm returns
the nonnegative vector $\tilde{x}_{opt} \in R^d$ that satisfies
\begin{equation} \label{eqn:tildexopt}
\tilde{x}_{opt} = \arg\min_{x \in \mathbb{R}^d, x \geq
0}\TNorm{\tilde{H}D\left(Ax - b\right)}^2.
\end{equation}
In section \ref{sec:optr} we will argue that, for any $\epsilon
\in (0, 1/3]$, if we set
\begin{equation}\label{eqn:r}
r \geq 684 c_o^2 (d+1) \log(n) \log(342c^2 (d+1) \log(n) /
\epsilon^2) / \epsilon^2,
\end{equation}
then $\TNorm{A\tilde{x}_{opt}-b}^2$ is at most $(1+\epsilon)$
worse than the true optimum $\TNorm{Ax_{opt}-b}^2$. This is a
sufficient (but not necessary) condition for $r$ in order to
satisfy the relative error guarantees of
equation~(\ref{eqn:result1_intro}). Indeed, in the experiments of
section \ref{sec:exp} we will argue that empirically a much
smaller value of $r$, for example $r = d + 20$, suffices.
\begin{algorithm}[h]
\begin{framed}

\textbf{Inputs:} $A \in \mathbb{R}^{n \times d}$, $b \in
\mathbb{R}^n$, positive integer $r < n$.

\vspace{0.1in}

\textbf{Output:} a nonnegative vector $\tilde{x}_{opt} \in
\mathbb{R}^d$.

\begin{enumerate}

\item Let $H_n$ be the $n \times n$ normalized Hadamard-Walsh matrix.

\item Let $S$ be an $n \times n$ diagonal matrix such that for all $i \in [n]$,
    $$
    S_{ii}
            = \left\{ \begin{array}{ll}
                         \sqrt{n/r}, & \mbox{with probability $r/n$} \\
                         0, & \mbox{otherwise}
                      \end{array}
              \right.
    $$
    %

\item Let $\tilde{H}$ be the matrix consisting of the non-zero rows of $ S H_n$. \\
(Notice that $\tilde{H}$ has -- in expectation -- $r$ rows.)

\item Construct the $n \times n$ diagonal matrix $D$ such that, for all $i\in [n]$, \\ $D_{ii} =
+1$ with probability $1/2$; otherwise $D_{ii} = -1$.

\item Solve
$$\tilde{x}_{opt} = \arg\min_{x \in \mathbb{R}^d, x\geq 0}\TNorm{\tilde{H}DAx -\tilde{H}Db}^2,$$
using any standard NNLS solver and return the vector
$\tilde{x}_{opt}$.
\end{enumerate}

\end{framed}
\caption{The \textsc{RandomizedNNLS} algorithm.}
\label{alg:alg_sample_fast}
\end{algorithm}

\subsection{The proof of Theorem \ref{thm:main_result}}

\subsubsection{A random projection type result}
\label{sxn:subspacesampling}

In this section we prove a random projection type result based on the
so-called subspace sampling procedure~\cite{DMM07} that might be
of independent interest. In particular, given a matrix $\Phi \in
\mathbb{R}^{n \times d}$ with $n \gg d$ (a.k.a., $\Phi$ is tall
and thin), and \textit{any} vector $y \in \mathbb{R}^d$, we argue
that the $\ell_2$ norm of $\Phi y$ can be estimated very
accurately by $\tilde{\Phi}y$, where $\tilde{\Phi}$ consists of
small subset of (appropriately rescaled) rows of $\Phi$.

More specifically, consider the \textsc{SubspaceSampling}
algorithm described below. This algorithm selects a small subset
of rows of $\Phi$ to construct $\tilde{\Phi}$; notice that
$\tilde{\Phi}$ has -- in expectation -- at most $r$ rows. Also
notice that $\tilde{\Phi}$ contains the $i$-th row of $\Phi$
(appropriately rescaled) if and only if $S_{ii}$ is non-zero.
Lemma \ref{lemma:subspace-sampling} bounds the approximation error
for our subspace sampling algorithm.

\begin{algorithm}[h]
\begin{framed}

\textbf{Input:} $\Phi \in \mathbb{R}^{n \times d}$, integer $r <
n$, set of probabilities $p_i\geq 0$, $i \in [n]$ s.t. $\sum_{i
\in [n]} p_i = 1$.

\textbf{Output:} $\tilde{\Phi} \in \mathbb{R}^{\tilde{r} \times
d}$, with $\Exp\left(\tilde{r}\right) \leq r$.

\vspace{0.1in}

\begin{enumerate}
    \item Let $S$ be the $n \times n$ diagonal matrix such that for all $i \in [n]$,
    $$
    S_{ii}
            = \left\{ \begin{array}{ll}
                         1/\sqrt{\min\{1,rp_i\}} & \mbox{,with probability $\min\{1,rp_i\}$} \\
                         0 & \mbox{,otherwise}
                      \end{array}
              \right.
    $$
    \item Let $\tilde{\Phi}$ be the matrix consisting of the non-zero rows of
    $S\Phi$. \\ (Notice that $\tilde{\Phi}$ has - in expectation - $r$ rows.)
\end{enumerate}
\end{framed}
\caption{\textsc{SubspaceSampling} algorithm} \label{alg:alg2}
\end{algorithm}

\begin{lemma}
\label{lemma:subspace-sampling} Let $\epsilon \in (0,1]$. Let
$\Phi$ be an $n \times d$ matrix ($n \gg d$), $U_{\Phi}$ be the $n
\times d$ matrix containing the left singular vectors of $\Phi$,
and $U_{\Phi(i)}$ denote the $i$-th row of $U_{\Phi}$. Let
$\tilde{\Phi}$ be constructed using the \textsc{SubspaceSampling}
algorithm with inputs $\Phi$, $r$, and sampling probabilities
$p_i$. If for all $i \in [n]$,
\begin{equation}
\label{eqn:p} p_i \geq \beta \TNormS{U_{\Phi(i)}}/d,
\end{equation}
for some $\beta \in (0,1]$, and the parameter $r$ satisfies
\begin{equation}
\label{eqn:r} \frac{r}{\log(r)} \geq \frac{ 9c_o^2 d}{\beta
\epsilon^2},
\end{equation}
for a sufficiently large constant $c_o$, then with probability at
least $2/3$ all $d$-dimensional vectors $y$ satisfy,
\begin{equation}
\label{eqn:ally} \abs{\TNormS{\Phi y} - \TNormS{\tilde{\Phi} y}}
\leq \epsilon \TNormS{\Phi y}.
\end{equation}
\end{lemma}
\begin{Proof}
Let $\Phi = U_{\Phi} \Sigma_{\Phi} V_{\Phi}^T$ be the SVD of
$\Phi$ with $U_{\Phi} \in \mathbb{R}^{n \times d}$, $\Sigma_{\Phi}
\in \mathbb{R}^{d \times d}$, and $V_{\Phi} \in \mathbb{R}^{d
\times d}$. Let $S$ be the $n \times n$ diagonal matrix
constructed at the first step of algorithm
\textsc{SubspaceSampling}, and let $U_{\Phi}^T S^TS U_{\Phi} = I +
E$, where $I$ is the $d \times d$ identity matrix, and $E$ some $d
\times d$ matrix. Then, using these two definitions,
submultiplicativity, and the orthogonality and unitary invariance
of $U_{\Phi}$ and $V_{\Phi}$,
\begin{eqnarray*}
\label{eqn:mainlemma:eq1}
\abs{\TNormS{\Phi y} - \TNormS{\tilde{\Phi} y}} &=&  \abs{ y^T \Phi^T \Phi y - y^T \Phi^T S^TS \Phi y}\\
\label{eqn:mainlemma:eq2}
         &=&   \abs{ y^T V_{\Phi} \Sigma^{2}_{\Phi} V_{\Phi}^T y - y^T V_{\Phi} \Sigma_{\Phi} U_{\Phi}^T S^TS U_{\Phi} \Sigma_{\Phi} V_{\Phi}^T y}\\
\label{eqn:mainlemma:eq3}
         &=&  \abs{ y^T V_{\Phi} \Sigma^{2}_{\Phi} V_{\Phi}^T y - y^T V_{\Phi} \Sigma_{\Phi} (I+E) \Sigma_{\Phi} V_{\Phi}^T y}  \\
\label{eqn:mainlemma:eq5}
         &=&  \abs{y^T V_{\Phi} \Sigma_{\Phi} E \Sigma_{\Phi} V_{\Phi}^T y}  \\
\label{eqn:mainlemma:eq6}
         &\leq&  \TNorm{y^T V_{\Phi} \Sigma_{\Phi}} \TNorm{E}\TNorm{ \Sigma_{\Phi} V_{\Phi}^T y}  \\
\label{eqn:mainlemma:eq7}
         &=&  \TNorm{E}  \TNormS{\Sigma_{\Phi} V_{\Phi}^T y}  \\
\label{eqn:mainlemma:eq8}
         &=&  \TNorm{E}  \TNormS{U_{\Phi}\Sigma_{\Phi} V_{\Phi}^T y}  \\
\label{eqn:mainlemma:eq9}
         &=&  \TNorm{E} \TNormS{\Phi y}.
\label{eqn:mainlemma:eq10}
\end{eqnarray*}
Using Theorem 7 of \cite{DMM07} (originally proven by Rudelson and
Virshynin in~\cite{RV06}), we see that
\begin{equation}
\Exp\left({\TNorm{E}}\right) \leq c_o \sqrt{\frac{\log(r)}{\beta
r}}\TNorm{U_{\Phi}}\FNorm{U_{\Phi}} = c_o
\sqrt{\frac{d\log(r)}{\beta r}},
\end{equation}
for a sufficiently large constant $c_o$ ($c_o$ is not specified
in~\cite{RV06}). Markov's inequality implies that
\begin{equation}
\TNorm{E} \leq 3c_o \sqrt{\frac{d\log(r)}{\beta r}},
\end{equation}
with probability at least 2/3. Finally, using
equation~(\ref{eqn:r}) concludes the proof of the lemma.
%
\end{Proof}
\clearpage
\subsubsection{Another useful result}
Results in~\cite{AC06} imply the following lemma.
\begin{lemma}
\label{lem:HU} Let $U$ be an $n \times d$ orthogonal matrix ($n
\geq 20$ and $n \geq d$). Then, for all $i \in [n]$,
\begin{eqnarray}
\label{eqn:lem:HU_eqn2} \TNormS{\left(H_nDU\right)_{(i)}}
   &\leq& \frac{4.2 d \log n}{n},
\end{eqnarray}
holds with probability at least 0.9.
\end{lemma}

\subsubsection{The proof of Theorem \ref{thm:main_result}}
We are now ready to prove Theorem \ref{thm:main_result}. We apply
lemma \ref{lemma:subspace-sampling} for
$\Phi = \left[%
\begin{array}{cc}
  H_nDA & -H_nDb \\
\end{array}%
\right] \in R^{n \times (d+1)}$, the parameter $r$ of Theorem
\ref{thm:main_result}, sampling probabilities $p_i = 1/n$, for all
$i \in [n]$, $\beta = 1/\left(4.2 \log n\right)$, and $\epsilon' =
\epsilon / 3 \in (0,1/3]$, where $\epsilon \in (0,1]$ is the
parameter of Theorem \ref{thm:main_result}. Let $U_{\Phi}$ be the
$n \times (d+1)$ matrix of the left singular vectors of $\Phi$.
Note that $U_{\Phi}$ is exactly equal to $U_{\Phi} = H_n D U_{[A
\hspace{0.05in} -b]}$, where $U_{[A \hspace{0.05in} -b]}$ is the
$n \times (d+1)$ matrix of the left singular vectors of $[A
\hspace{0.05in} -b]$. Lemma \ref{lem:HU} for $U_{[A
\hspace{0.05in} -b]}$ and our choice of $\beta$, guarantee that
for all $i \in [n]$, with probability at least $0.9$
$$ 1/n \geq \beta \TNormS{\left(H_nDU_{[A - b]}\right)_{(i)}}/d \Rightarrow 1/n \geq \beta \TNormS{\left(U_{\Phi}\right)_{(i)}}/d.$$
The latter inequality implies that assumption~(\ref{eqn:p}) of
Lemma \ref{lemma:subspace-sampling} holds. Also, our choice of
$r$, which satisfies inequality~(\ref{eqn:result0_intro}) in
Theorem \ref{thm:main_result}, our choice of $\beta$, and our
choice of $\epsilon$, guarantee that assumption~(\ref{eqn:r}) of
Lemma \ref{lemma:subspace-sampling}
is also true. Since all $d$-dimensional vectors $y$ satisfy  equation~(\ref{eqn:ally}), we pick $y=\left[%
\begin{array}{c}
  x_{opt} \\
  1 \\
\end{array}%
\right]$ and $\tilde{y}=\left[%
\begin{array}{c}
  \tilde{x}_{opt} \\
  1 \\
\end{array}\right],$ thus getting that with probability at least 2/3:
\begin{equation}
\label{eqn:rel1} (1-\epsilon') \TNormS{H_n D A x_{opt}-H_n Db }
\leq \TNormS{ \tilde{H}_n D A x_{opt}-\tilde{H}_n Db} \leq
(1+\epsilon')\TNormS{H_n D A x_{opt} -H_n Db},
\end{equation}
and
\begin{equation}
\label{eqn:rel2}
(1-\epsilon') \TNormS{ H_n D A \tilde{x}_{opt} -H_n Db} \leq
\TNormS{ \tilde{H}_n D A \tilde{x}_{opt}-\tilde{H}_n Db} \leq
(1+\epsilon')\TNormS{H_n D A \tilde{x}_{opt} -H_n Db}.
\end{equation}
Manipulating equations~(\ref{eqn:rel1}) and (\ref{eqn:rel2}) we
get
\begin{eqnarray*}
\TNormS{ H_n D A \tilde{x}_{opt} -H_n Db} &\leq& \frac{1}{1-\epsilon'} \TNormS{ \tilde{H}_n DA \tilde{x}_{opt}-\tilde{H}_n Db} \\
&\leq&   \frac{1}{1-\epsilon'} \TNormS{\tilde{H}_n DA x_{opt}-\tilde{H}_n Db }  \\
&\leq&   \frac{1+\epsilon'}{1-\epsilon'} \TNormS{H_n DA x_{opt}-H_n Db }  \\
&\leq&   \left(1 + 3\epsilon'\right) \TNormS {H_n DA x_{opt}-H_n
Db} \\
&\leq&   \left(1 + \epsilon\right) \TNormS {H_n DA x_{opt}-H_n
Db}.
\end{eqnarray*}
The second inequality follows since $\tilde{x}_{opt}$ is the
optimal solution of the NNLS problem of eqn.
(\ref{eqn:tildexopt}), thus $x_{opt}$ is a sub-optimal solution,
and the fourth inequality follows since
$\left(1+\epsilon'\right)/\left(1-\epsilon'\right) \leq
1+3\epsilon'$, for all $\epsilon' \in (0,1/3]$. In the last
inequality, we set $\epsilon' = \epsilon /3$. To conclude the
proof, notice that $H_n D$ is an orthonormal square matrix and can
be dropped without changing a unitarilly invariant norm. Finally,
since Lemmas \ref{lemma:subspace-sampling} and \ref{lem:HU} fail
with probability at most $1/3$ and $1/10$ respectively, the union
bound implies that Theorem \ref{thm:main_result} fails with
probability at most $0.5$.

\subsection{What is the minimal value of $r$ ?}
\label{sec:optr} To derive values of $r$ for which the
\textsc{RandomizedNNLS} algorithm satisfies the relative error
guarantees of Theorem \ref{thm:main_result}, we need to solve
equation~(\ref{eqn:result0_intro}); this is hard since the
solution depends on the Lambart $W$ function. Thus, we identify a
range of values of $r$ that are sufficient for our purposes. Using
the fact that for any $\alpha \geq 4$, and for any $\gamma \geq 2
\alpha \log(\alpha)$,
$$ \frac{\gamma}{\log(\gamma)} \geq \alpha ,$$
and by setting $\alpha = 342 c_o^2 (d+1) \log(n) / \epsilon^2 $ in
equation~(\ref{eqn:result0_intro}) (note that $342 c_o^2 (d+1)
\log(n) / \epsilon^2 \geq 4$), it can be proved that every $r$
such that
\begin{equation}
\label{eqn:result2_intro} r \geq 684 c_o^2 (d+1) \log(n) \log(342
c_o^2 (d+1) \log(n) / \epsilon^2) / \epsilon^2,
\end{equation}
satisfies the inequality \ref{eqn:result0_intro} ($c_o$ is the
constant of Theorem \ref{thm:main_result}).

\subsection{Running time analysis}
\label{sec:time} In this subsection we analyze the running time of
our algorithm. Let $r$ be the minimal value that satisfies
equation (\ref{eqn:result2_intro}). First, computing $DA$ and $Db$
takes $O(nd)$ time. Since $\tilde{H}$ has in expectation $r$ rows,
Ailon and Liberty in~\cite{AL08} argue that the computation of
$\tilde{H}DA$ and $\tilde{H}Db$ takes $O(nd\log r)$ time. For our
choice of $r$, this is
$$T_{precond} = O(nd \log ( d \log(n) / \epsilon^2 )).$$ After
this preconditioning step, we employ an NNLS solver on the smaller
problem. The computational cost of the NNLS solver on the small
problem was denoted as $T_{NNLS}(r,d)$ in Theorem
\ref{thm:main_result}. $T_{NNLS}(r,d)$ cannot be specified exactly
since theoretical running times for exact NNLS solvers are
unknown. In the sequel we comment on the computational costs of
some well defined segments of some NNLS solvers.

The NNLS formulation of Definition \ref{def:NNLS} is a convex
quadratic program, and is equivalent to
\begin{eqnarray*}
\min_{x \in \mathbb{R}^d, x\geq 0} x^T Q x - 2q^T x,  \\
\end{eqnarray*}
where $Q = A^T A \in \mathbb{R}^{d\times d}$ and $q = A^Tb \in
\mathbb{R}^d$. Computing $Q$ and $q$ takes $O(nd^2)$ time, and
then the time required to solve the above formulation of the NNLS
problem is independent of $n$. Using this formulation, our
algorithm would necessitate $T_{precond}$ time for the computation
of $\tilde{H}DA$ (the preconditioning step described above), and
then $\tilde{Q} = \left(\tilde{H}DA\right)^T\tilde{H}DA$ and
$\tilde{q} = \left(\tilde{H}DA\right)^Tb$ can be computed in
$T_{MM} = O(rd^2)$ time; given our choice of $r$, this implies
$$ T_{MM} = O(d^3 \log(n) / \epsilon^2).$$ Overall, the standard
approach would take $O(nd^2)$ time to compute $Q$, whereas our
method would need only $T_{precond}$ + $T_{MM}$ time for the
construction of $\tilde{Q}$. Note, for example, that when
$n=O(d^2)$ and regarding $\epsilon$ as a constant, $\tilde{Q}$ can
be computed $O(d / \log(d))$ times faster than $Q$.

On the other hand, many standard implementations of NNLS solvers
(and in particular those that are based on active set methods)
work directly on the formulation of Definition \ref{def:NNLS}. A
typical cost of these implementations is of the order $O(nd^2)$
per iteration. Other approaches, for example the NNLS method of
\cite{KSD07}, proceed by computing matrix-vector products of the
form $Au$, for an appropriate $d$-dimensional vector $u$, thus
cost typically $O(nd)$ time per iteration. In these cases our
algorithm needs again $T_{precond}$ preprocessing time, but costs
only $O(rd^2)$ or $O(rd)$ time per iteration, respectively. Again,
if given our choice of $r$, the computational savings \emph{per
iteration} are comparable with the $O(d / \log(d))$ speedup
described above.

\section{Experimental Evaluation}
\label{sec:exp}

In this section, we experimentally evaluate our
\textsc{RandomizedNNLS} algorithm on \textbf{(i)} large, sparse
matrices from a text-mining application, and \textbf{(ii)}
random matrices with varying sparsity. We mainly focus on
employing the state-of-the-art solver of \cite{KSD07} to solve the small NNLS problem.

\subsection{The TechTC300 dataset}
\label{sec:exppart1}

Our data come from the Open Directory Project (ODP)~\cite{odp}, a
multilingual open content directory of WWW links that is
constructed and maintained by a community of volunteer editors.
ODP uses a hierarchical ontology scheme for organizing site
listings. Listings on similar topics are grouped into categories,
which can then include smaller subcategories. Gabrilovich and
Markovitch constructed a benchmark set of 300 term-document
matrices from ODP, called TechTC300 (Technion Repository of Text
Categorization Datasets~\cite{GM04}), which they made publicly
available. Each term-document matrix of the TechTC300 dataset
consists of a total of 150 to 400 documents from two different ODP
categories, and a total of 15,000 to 35,000 terms. We chose this
dataset because we believe that it does represent an important
application area, namely text mining, and we do believe that the
results from our experiments will be representative of the
potential usefulness of our randomized NNLS algorithm in large,
sparse, term-document NNLS problems.

We present average results from 3,000 NNLS problems. More
specifically, for each of the 300 matrices of the TechTC300
dataset, we randomly choose a column from the term-document matrix
as the vector $b$, we assign the remaining columns of the same
term-document matrix to the matrix $A$, and solve the resulting
NNLS problem with inputs $A$ and $b$. We repeat this process ten
times for each term-document matrix of the TechTC300 dataset, and
thus solve a total of 3,000 problems. Whenever an NNLS routine is
called, it is initialized with the all-zeros vector. We evaluate
the accuracy and the running time of our algorithm when compared
to two standard NNLS algorithms. The first one is described
in~\cite{KSD07}\footnote{We would like to thank the authors
of~\cite{KSD07} for providing us with a Matlab
implementation of their algorithm.}, and the second
one is the active set method of \cite{LH74}, implemented as the
built-in function \emph{lsqnonneg} in Matlab. We would also like
to emphasize that in~\cite{KSD07} the authors compare their
approach to other NNLS approaches and conclude that their
algorithm is significantly faster. Note that the method of
\cite{KSD07} operates on the quadratic programming formulation
discussed in Section \ref{sec:RNNLS}\footnote{The actual
implementation involves computations of the form $t=Au$ and
$s=A^Tt$, avoiding the computation and storage of the matrix
$A^TA$.}, while \emph{lsqnonneg} operates on the formulation of
Definition \ref{def:NNLS}. Finally, we implemented our
\textsc{RandomizedNNLS} algorithm in Matlab. The platform used for
the experiments was a 2.0 GHz Pentium IV with 1GB RAM.

\begin{figure}
\label{fig:results}
    \begin{center}
    \includegraphics[scale=0.52,angle=0]{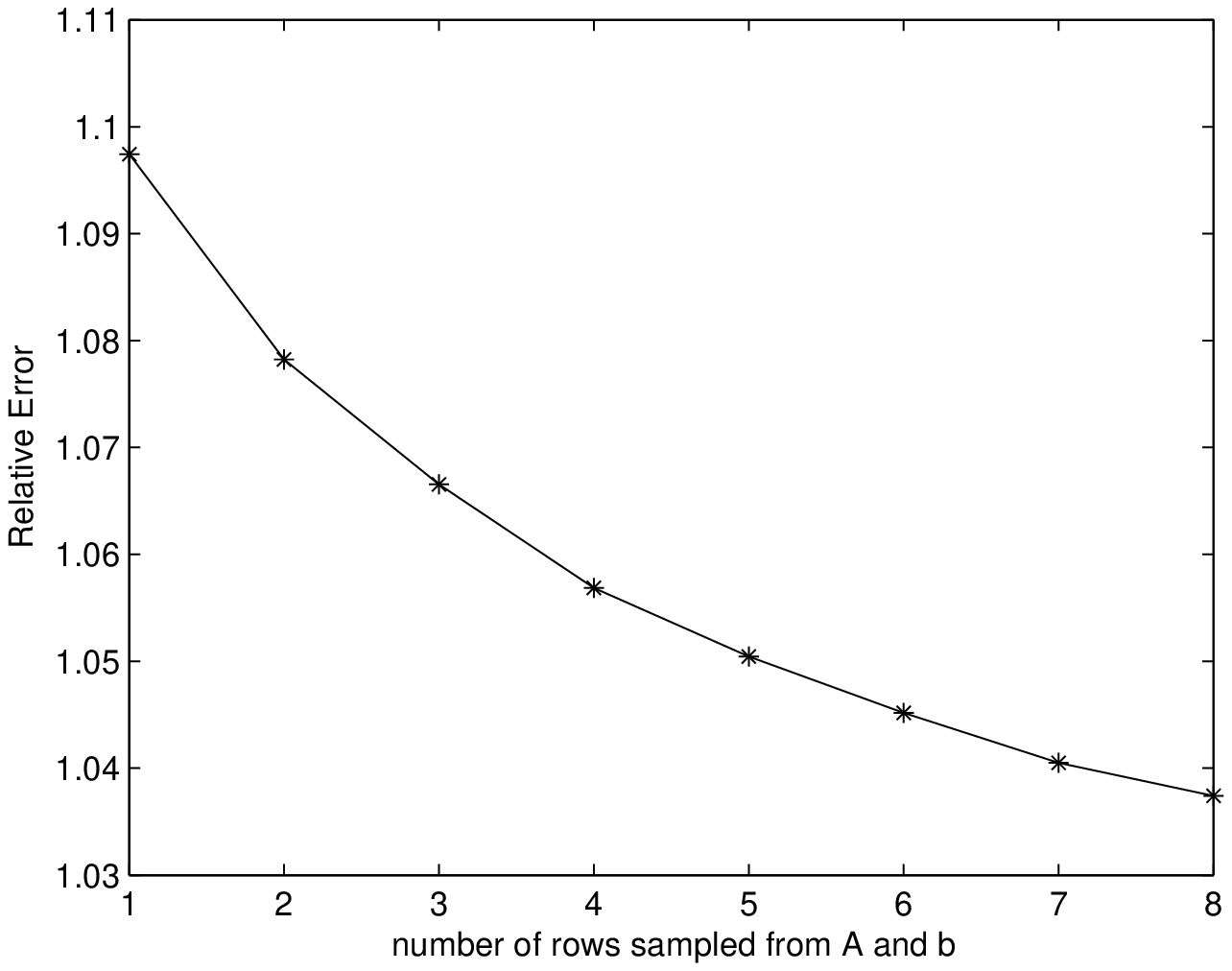}
    \includegraphics[scale=0.52,angle=0]{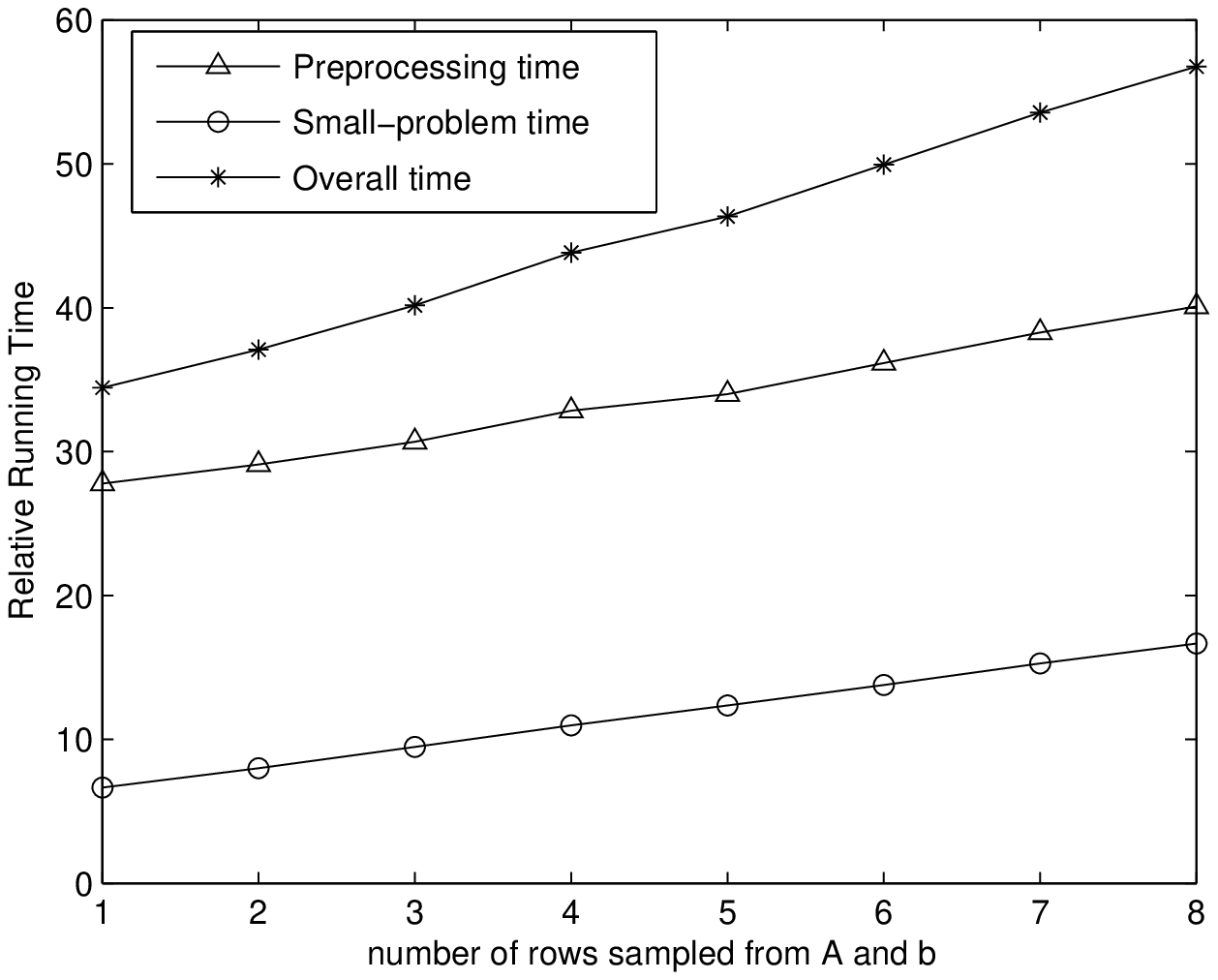}
    \end{center}
    \caption{
    Average results of the \textsc{RandomizedNNLS}
    algorithm compared to the algorithm of \cite{KSD07} on 3000 NNLS
    problems constructed from the TechTC300 dataset.
    \emph{Relative Error}$:= \TNorm{ A\tilde{x}_{opt}-b} / \TNorm{Ax_{opt}-b}$, while \emph{Overall Time}$:= 100 T_{\tilde{x}_{opt}} /
    T_{x_{opt}}$. $x_{opt}$ is computed with the method of \cite{KSD07} in
    $T_{x_{opt}}$ time, and $\tilde{x}_{opt}$  with the \textsc{RandomizedNNLS} algorithm (the last step employs the method of \cite{KSD07}) in
    $T_{\tilde{x}_{opt}}$ time. Points one through eight on the $x$-axis of the plots
    correspond to values of the parameter $r = d + i\cdot 50$ for $i=1\ldots
    8$, where $d$ is the number of columns of $A$.
    On the right panel, \emph{Preprocessing
    time} stands for the cost of the multiplication of $A$ and $b$ with $\tilde{H}D$ (multiplied by $100$ and divided by $T_{x_{opt}}$), and \emph{Small-problem time} stands
    for the cost of solving the small problem using the algorithm
    of \cite{KSD07} (multiplied by $100$ and divided by $T_{x_{opt}}$). For each point of the $x$-axis: \emph{Overall time} = \emph{Preprocessing
    time} +  \emph{Small-problem time}.}\label{fig:fig1}
\end{figure}

Our (average) results are shown in Figure \ref{fig:fig1}. We only
focus on the algorithm of~\cite{KSD07}, which was significantly
faster, running (on average) in five seconds, compared to more
than one minute for the \emph{lsqnonneg} function. We experimented
with eight different values of the parameter $r$, which dictates
the size of the small subproblem (see the \textsc{RandomizedNNLS}
algorithm). More specifically, we set $r$ to $d+i\cdot 50$, for
$i=1\ldots 8$, where $d$ is the number of columns in the matrix
$A$. Our results verify that ($i$) the \textsc{RandomizedNNLS}
algorithm is very accurate, ($ii$) that it reduces the running
time of the NNLS method of~\cite{KSD07}, and $(iii)$ that there
exists a natural tradeoff between the approximation accuracy and
the number of sampled rows. Notice, for example, that the running
time of the state-of-the-art NNLS solver of~\cite{KSD07} can be
reduced from two to three times, while the residual error is from
$4\%$ up to $10\%$ worse than the optimal residual error.

We briefly comment on the performance of \textsc{RandomizedNNLS}
when compared to the \emph{lsqnonneg} algorithm. As expected, the
accuracy results are essentially identical with the method
of~\cite{KSD07}, since both methods solve the NNLS problem
exactly. Our speedup, however, was much more significant, ranging
from 14-fold to 10-fold for $r = d+50$ and $r = d+400$
respectively (data not shown).

\subsection{Sparse vs dense NNLS problems}
\label{sec:sparsedense}

The astute reader might notice that we evaluated the performance of our
algorithm in a rather adversarial setting. The TechTC300 data are
quite sparse, hence existing NNLS methods would operate on sparse
matrices. However, our preprocessing step in the
\textsc{RandomizedNNLS} algorithm destroys the sparsity, and the
induced subproblem becomes dense. Thus, we are essentially
comparing the time required to solve a sparse, large NNLS problem
to the time required to solve a dense, small NNLS problem. If the
original problem were dense as well, we would expect more
pronounced computational savings. In this section we experiment
with random matrices of varying density in order to confirm this hypothesis.

First, it is worth noting that the sparsity of the input matrix
$A$ and/or the target vector $b$ do not seem to affect the
approximation accuracy of the \textsc{RandomizedNNLS} algorithm.
This should not come as a surprise since our results in Theorem
\ref{thm:main_result} do not make any assumptions on the inputs
$A$ and $b$. Indeed, our experiments in Figure 2 confirm our
expectations.

Prior to discussing our experiments on random matrices of varying
density, it is worth noting that the NNLS solver of~\cite{KSD07}
has a running time that is a function of the number of non-zero
entries of $A$. Indeed, the method of~\cite{KSD07} is an iterative
method where the computational bottleneck in the $j$-th iteration
involves computations of the form $A^T A u$, for a $d$-dimensional
vector $u$. \cite{KSD07} implemented their algorithm by computing
the two matrix-vector products $Au$ and $A^T \left(Au\right)$
separately, thus never forming the matrix $A^T A$ and thus taking
advantage of the sparsity of $A$. Indeed, NNLS problems with
sparse coefficient matrices $A$ are solved faster than NNLS
problems with similar-size dense coefficient matrices $A$ by using
the method of \cite{KSD07}\footnote{The authors of \cite{KSD07}
performed extensive numerical experiments to verify that
observation; for example see the last row of Table 4 on page 14 in
\cite{KSD07} and notice that the running time of their method
increases as the density of $A$ increases.}.

In order to measure how the speedup of our \textsc{RandomizedNNLS}
algorithm improves as the matrix $A$ and vector $b$ become denser,
we designed the following experiment. First, let the
\textit{density} of an NNLS problem denote the percentage of
non-zero entries in $A$ and $b$; for example, $density(A,b) = 10
\%$ means that approximately $0.9(nd + n)$ entries in the $n
\times d$ matrix $A$ and the $n \times 1$ vector $b$ are zero. We
chose six density parameters ($2\%$, $4\%$, $8\%$, $16\%$, $32\%$,
and $64\%$) and generated $100$ NNLS problems for each density
parameter. More specifically, we first constructed ten $n \times
(d+1)$ random matrices with the target density (the non-zero
entries are normally distributed in $[0,1]$). Then, for each
matrix, we randomly selected one column to form the vector $b$ and
assigned the remaining $d$ columns to the matrix $A$. We repeated
this selection process ten times for each of the ten $n \times
(d+1)$ matrices, thus forming a set of $100$ NNLS problems with
inputs $A$ and $b$. We fixed the dimensions to $n = 10,000$ and
$d=300$ and we experimented with four values of $r=( d, d+50,
d+100, d+150)$. In Figure 2 we present average results over the
$100$ NNLS problems for each choice of the density parameter.
Notice that increasing the density of the inputs $A$ and $b$, the
computational gains increase as well. On top of that, our method
becomes more accurate while the number of the zero entries in $A$
and $b$ become fewer. Notice for example, on the right plot of
Figure 2, when $r=300$, the two extreme cases ($density=2\%$ and $
density = 64\%$) correspond to an $18\%$ and a $4\%$ loss in
accuracy, respectively. Given these two observations as well as
the actual times of $T_{x_{opt}}$ (see the caption of Figure 2),
we conclude that the random projection ideas empirically seem more
promising for dense rather than sparse NNLS problems.

\begin{figure}
\label{fig:results2}
    \begin{center}
    \includegraphics[scale=0.52,angle=0]{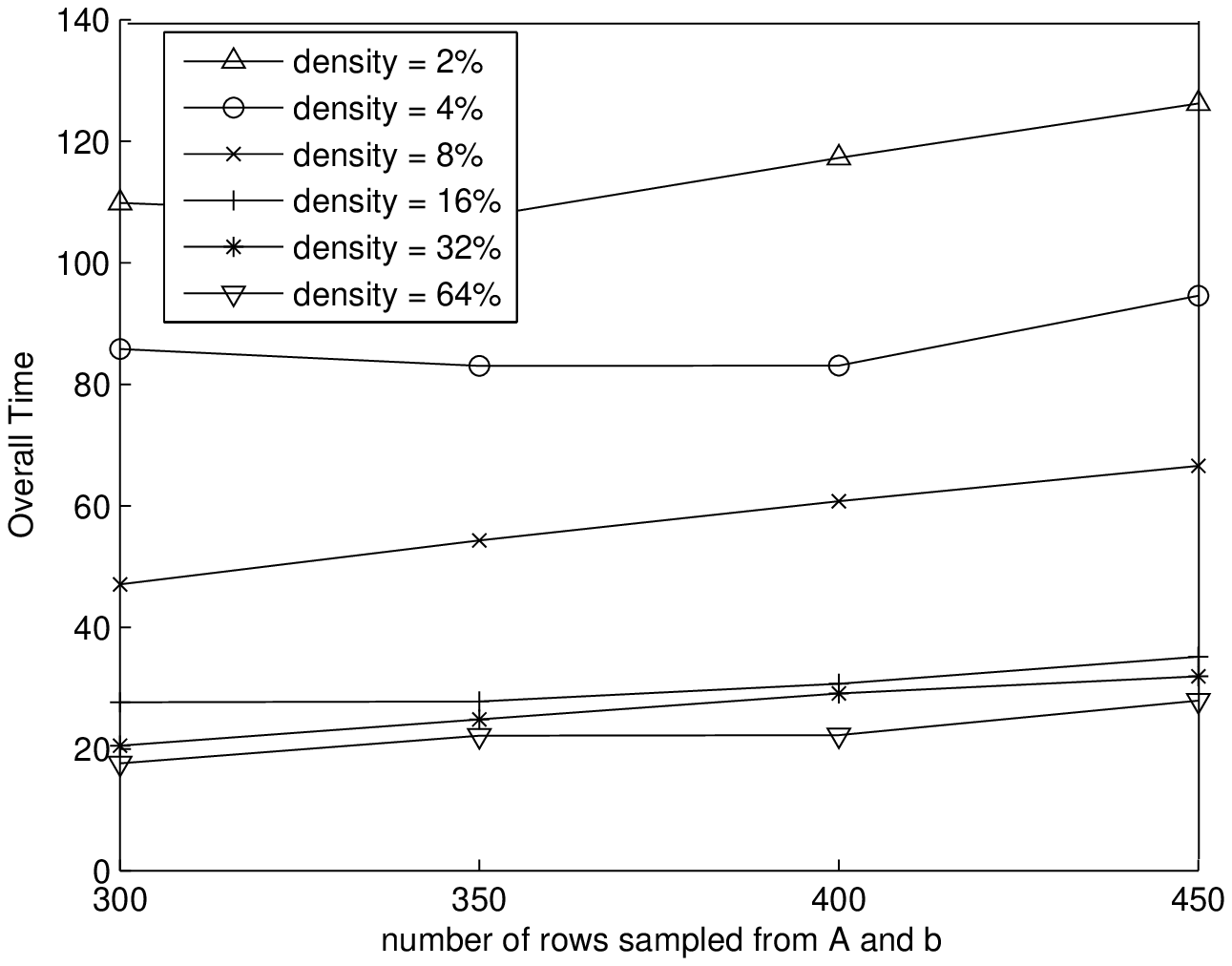}
    \includegraphics[scale=0.52,angle=0]{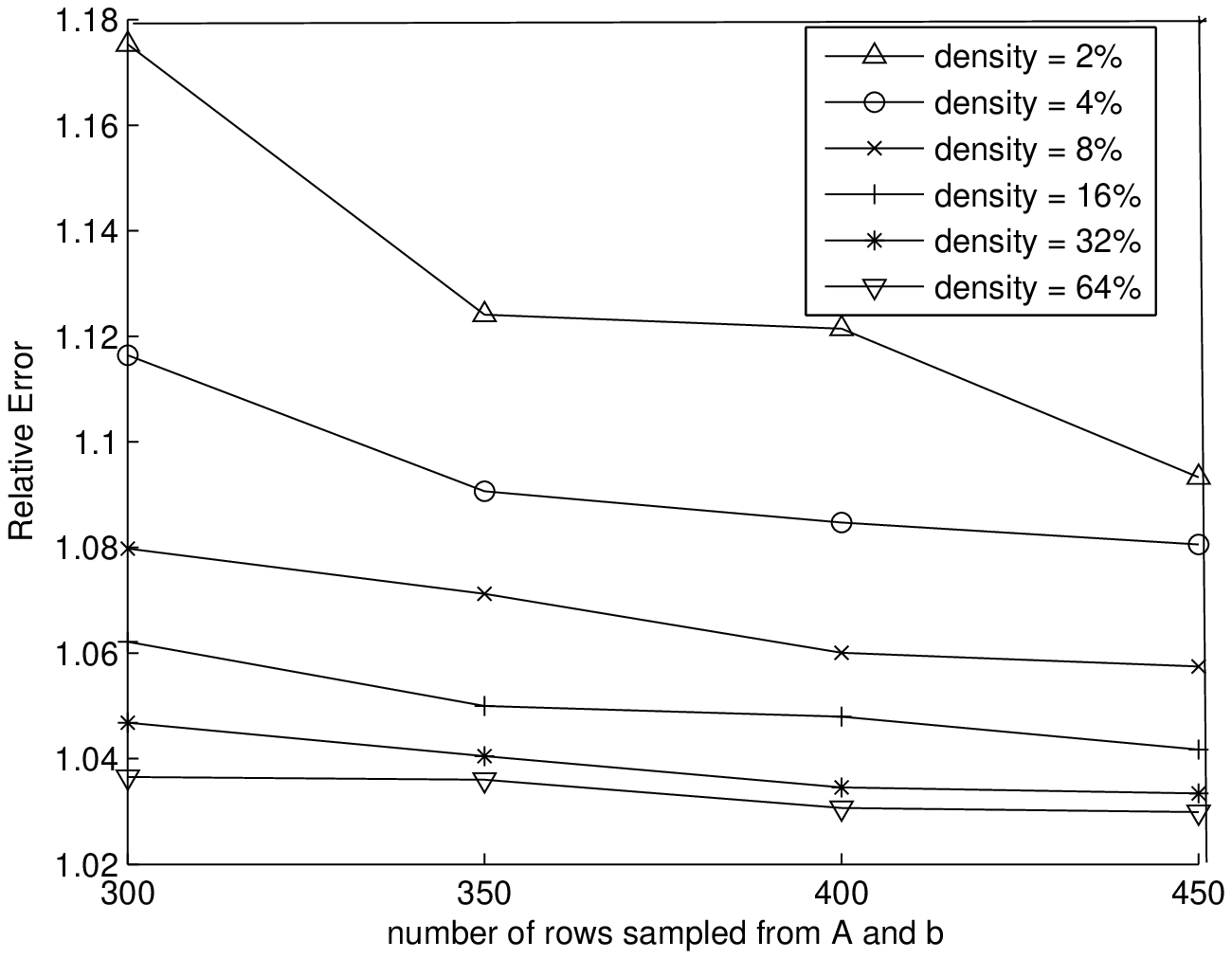}
    \end{center}
    \caption{
    Average results of the \textsc{RandomizedNNLS}
    algorithm compared to the algorithm of \cite{KSD07} on six sets of $100$ NNLS
    problems with different density.
    \emph{Relative Error}$:= \TNorm{ A\tilde{x}_{opt}-b} / \TNorm{Ax_{opt}-b}$, while \emph{Overall Time}$:= 100 T_{\tilde{x}_{opt}} /
    T_{x_{opt}}$. $x_{opt}$ is computed with the method of \cite{KSD07} in
    $T_{x_{opt}}$ time, and $\tilde{x}_{opt}$  with the \textsc{RandomizedNNLS} algorithm (the last step employs the method of \cite{KSD07}) in
    $T_{\tilde{x}_{opt}}$ time. For the six density parameters $(2\%, 4\%, 8\%, 16\%, 32\%,
    64\%)$, $T_{x_{opt}}$ was on average (0.26 sec, 0.40 sec, 0.94 sec,
    2.89 sec, 6.27 sec, 10.57 sec), respectively.    }
\end{figure}


\section{Conclusions}

We presented a random projection algorithm for the Nonnegative
Least Squares Problem. We experimentally evaluated our algorithm
on a large, text-mining dataset, and verified that, as promised in
our theoretical findings, practically it does give very accurate
approximate solutions, while outperforming two standard NNLS
methods in terms of computational efficiency. Future work includes
the extension of our theoretical findings of Theorem 1 to NNLS
problems with multiple right hand side vectors. An immediate
application of this would be the computation of Nonnegative Matrix
Factorizations based on Alternating Least Squares type
approaches~\cite{KP07a, KP08}. Finally, notice that, since our
analysis is independent of the type of constraints on the vector
$x$, our main algorithm can be employed to approximate a
least-squares problem with any type of constraints on $x$.

\paragraph{Acknowledgements:} We would like to thank Kristin P. Bennett
and Michael W. Mahoney for useful discussions. The first author
would like also thank the Institute of Pure and Applied
Mathematics of the University of California at Los Angeles for its
generous hospitality during the period Sept. 2008 - Dec. 2008,
when part of this work was done.

\bibliographystyle{abbrv}
\bibliography{RPI_BIB}

\end{document}

%% file: newcmds.tex
\newcommand{\FNorm }[1]{\mbox{}\left\|#1\right\|_F  }

\newcommand{\FNormS}[1]{\mbox{}\left\|#1\right\|_F^2}

\newcommand{\TNorm }[1]{\mbox{}\left\|#1\right\|_2  }
\newcommand{\TNormS}[1]{\mbox{}\left\|#1\right\|_2^2}

\newcommand{\setlinespacing}[1]%
           {\setlength{\baselineskip}{#1 \defbaselineskip}}

\newcommand{\abs }[1]{\left|#1\right|}

\newcommand{\BlockDiagk}[1]{\mbox{}\left(%
\begin{array}{cc}
  \Sigma_{k} & \bf{0} \\
  \bf{0} &  \Sigma_{\rho-k}\\
\end{array}\right)}

\newcommand{\BlockDiagkk}[1]{\mbox{}\left(%
\begin{array}{cc}
  \Sigma_{k} & \bf{0} \\
  \bf{0} & \bf{0} \\
\end{array}\right)}

\newcommand{\BlockDiagkrk}[1]{\mbox{}\left(%
\begin{array}{cc}
  \bf{0} & \bf{0} \\
  \bf{0} & \Sigma_{\rho-k} \\
\end{array}\right)}

\newcommand{\BlockDiagkkh}[1]{\mbox{}\left(%
\begin{array}{c}
  \Sigma_{k} \\
  \bf{0} \\
\end{array}\right)}

\newcommand{\BlockDiagkrkh}[1]{\mbox{}\left(%
\begin{array}{c}
  \bf{0} \\
  \Sigma_{\rho-k} \\
\end{array}\right)}

\newtheorem{definition}{Definition}

\newtheorem{lemma}{Lemma}
\newtheorem{theorem}{Theorem}

\newenvironment{Proof}{\noindent {\em Proof:}}{\\\hspace*{\fill}\mbox{$\diamond$}}

\long\def\killtext#1{}

\def\Exp{\hbox{\bf{E}}}

